\documentclass{article}
\usepackage{amsthm,amsmath,amssymb}
\usepackage{mathrsfs}
\usepackage{bm}
\usepackage{geometry}
\usepackage{graphicx}
\usepackage{subfigure}
\usepackage{caption,subcaption}
\usepackage[affil-it]{authblk}
\usepackage{cite}
\begin{document}
\title{Simulation of smooth models of potentials with singular point using Many-Interacting-Worlds Method}
\author{Wen Chen}
\author{An Min Wang \thanks{anmwang@ustc.edu.cn}}
\affil{Department of Modern Physics, University of Science and Technology of China}
\maketitle
\begin{abstract}
The deterministic many-interacting-worlds method proposed in 2014 showed potential among the numerous interpretation of quantum mechanics. The successful application of this method in harmonic oscillator has been promoted for a long time. In this article we continue the idea about using this method to solve some bounded systems different from harmonic oscillator potential and extend to 2 dimension cases. We focus on the potential with singularity like coulomb potential and finite trap potential by some asymptotic smooth method. The numerical simulation mainly based on the dynamical algorithm proposed in many-interacting-worlds method will be used to approach the stationary states of given systems. Our results shows the consistency to the matrix Numerov method in standard quantum mechanics in solving bounded systems and provides the possibility to solve more complex systems.
\end{abstract}
\section{Introduction}
Hall, Deckert, and Wiseman proposed a many-interacting-worlds (MIW) method through  a huge, but finite, number of classical ``worlds'' and quantum effects arise from the interaction between these worlds \cite{hall2014quantum}. Similar to the quantum potential in Bohm mechanics, the interworld potential for $N$ countable ``worlds'' can be written as 
\begin{equation}\label{1}
U_{N}(\mathbf{X})=\sum^{N}_{n=1}\sum^{K}_{k=1}\left.\frac{\hslash^{2}}{8m^{k}P(\mathbf{q})^{2}}\left(\frac{\partial P(\mathbf{q})}{\partial q^{k}}\right)^{2}\right|_{\mathbf{q}=\mathbf{x}_{n}},
\end{equation}
where $\mathbf{X}=\{\mathbf{x}_{1},\dots ,\mathbf{x}_{N}\}$  denotes the worlds configurations,  $\mathbf{x}_{n}=\{x^{1}_{n}(t),\dots ,x^{K}_{n}(t)\}$ denotes the world-particle position, $m^{k}$ is the mass of the $k$th particle, $P(\mathbf{q})$ is the probability density of system. MIW has been used in reproducing some typical quantum phenomena \cite{hall2014quantum}
, stationary states of one dimensional and multidimensional quantum harmonic oscillator \cite{ghadimi2018nonlocality}\cite{hall2014quantum}\cite{herrmann2017eigenstates}\cite{mckeague2016convergence}\cite{mckeague2016stein}\cite{mckeague2023stein}\cite{sturniolo2018computational}, particles with spin 1/2 \cite{lombardini2024interacting}. 

In practice there are two different simulation ways in these models. One way is constructing proper sequences of worlds configurations with approximation of probability density function by solving the minimum problem of Hamiltonian. Most obtain the approximation of density based on stein's method and derive the interworld potential for given state. Analytically solving the equalities when Hamiltonian takes minimum can obtain the distribution of worlds configurations and the minimum of Hamiltonian will converge to the exact eigenenergy of given state when the number of worlds tends to infinity. Another way is the MIW dynamical algorithm by integrating the equations of motion using velocity Verlet algorithm. The energy reduces to local minimum with proper steps and large iterations. Worlds configurations are stationary when the forces constituted by classical force and quantum force are balanced. In practice MIW dynamical algorithm performed well in smooth potential with local minimum especially like harmonic oscillator. The hypothesis ``no crossing'' of trajectories and worlds is guaranteed by proper approximation of quantum potential and force based on proper approximation of density. In evolution proper time interval avoids worlds getting too close. All of these work mainly focus on the harmonic oscillator model for the smoothness of its potential and density. 

When it comes to some potentials without local minimum or not smooth enough at some points (potentials with singular point), the MIW dynamical algorithm may fail in evolution to stationary states. In \cite{chen2023extension} we first consider the one dimensional Coulomb potential model using the analytical solution method and obtain a right results which inspire us to explore the possibility of application of dynamical MIW algorithm in multidimensional Coulomb models. In this article we focus on some special potentials using MIW dynamical algorithm. We consider some smooth models to deal with potential without local minimum like Coulomb potential which make the MIW dynamic possible in such potential. We also consider the smooth model for finite depth trap potential that is not smooth enough. The method of approximation of density we used is kernel estimation of density which is proved a good choice in MIW method. The MIW dynamical algorithm perform successfully in simulation of ground states of these models. Besides the simulation of ground states, we firstly apply the MIW dynamical algorithm to excited state of 2-dimensional system and the results show the consistency with traditional matrix Numerov method. Our work provides the possibility to consider the MIW method in higher dimensional systems and multi-particle systems.
\section{Kernel estimation of density}
The approximation of density is very important in MIW for the equally weighted worlds. The countable worlds are involved in MIW by different forms of empirical distribution. In most work based on stein's method, the approximation of density takes the form $P_{N}(\mathbf{x}_{n};\mathbf{X})\approx P_{N}(\mathbf{x}_{n};\{\mathbf{x}_{n},\mathbf{x}_{n+1}\})\approx P_{N}(\mathbf{x}_{n};\{\mathbf{x}_{n},\mathbf{x}_{n-1}\})$ which is piecewise and discrete. For the convenience of further computation, the approximation of derivation of density in interworld potential and force restricts the number of worlds in $P_{N}$ which means the best choice to simplify the computation is containing the world $\mathbf{x}_{n}$ and its nearest world $\mathbf{x}_{n+1}$ or $\mathbf{x}_{n-1}$. Especially in MIW dynamical algorithm the quantum force will be more complex if we try to smooth the approximation by involving more neighbor worlds, which makes it easier to crash when the worlds evolve too close due to some numerical errors where the trajectory crossing is forbidden. Only two nearest worlds in density estimator leads to five worlds involved in quantum force of one dimensional example. In multidimensional cases too close worlds require the selection of time interval more cautious and the evolution will be unstable if the time interval too large.

A more smooth method to estimate the density is using kernel function. The best choice is the sample-point adaptive estimator \cite{scott2015multivariate}
\begin{equation}\label{2}
f(\mathbf{x};\mathbf{X},\{h(\mathbf{x}_{n})\})=\frac{1}{N}\sum_{n=1}^{N}\frac{1}{h(\mathbf{x}_{n})^{d}}K\left(\frac{\mathbf{x}-\mathbf{x}_{n}}{h(\mathbf{x}_{n})}\right)
\end{equation}
where $d$ denotes the dimensions of $\mathbf{x}$. $K(\mathbf{x})$ is the kernel function and $h(\mathbf{x}_{n})$ is the bandwidth dependent on sample points $\mathbf{X}=\{\mathbf{x}_{1},\dots ,\mathbf{x}_{N}\}$ which is worlds configurations in MIW. The Gaussian kernel and exponential kernel have been tested in harmonic oscillator models with convergent results obtained\cite{hall2014quantum}\cite{herrmann2017eigenstates}\cite{mckeague2016convergence}\cite{mckeague2016stein}\cite{mckeague2023stein}\cite{sturniolo2018computational}. Different from the discrete approximation of the derivation of interworld potential, if we consider the original expression of quantum potential for single particle in Bohmian mechanics
\begin{equation}\label{3}
Q(\mathbf{x}_{n})=-\left.\frac{\hbar^{2}}{2m}\frac{\nabla^{2}\sqrt{P(\mathbf{q})}}{\sqrt{P(\mathbf{q})}}\right|_{\mathbf{q}=\mathbf{x}_{n}}
\end{equation}
at $\mathbf{x}_{n}$ and replace $P(\mathbf{q})$ with the kernel estimator $P_{N}(\mathbf{q};\mathbf{X})\equiv f(\mathbf{q};\mathbf{X},\{h(\mathbf{x}_{n})\})$, the approximation of quantum force is unnecessary. A smooth unimodal kernel that symmetric about the origin is recommended in \cite{scott2015multivariate}
. According to the relative efficiencies table in \cite{scott2015multivariate}
, most kernels with good efficiencies are not differentiable at certain points because of the factor $I_{[-1,1]}$. Though Gaussian kernel is not the most efficient kernel, the smoothness in calculation makes it candidate. We will also test some other kernels in later simulation.

In general the bandwidth $h$ can be determined by minimizing the mean squared error (MSE) of the form
\begin{equation}\label{4}
\text{MSE}\{f(\mathbf{x})\}=\text{E}[f(\mathbf{x})-P(\mathbf{x})]^{2}.
\end{equation}
Follow the Theorem 6.3 in \cite{scott2015multivariate}
, we can obtain the initial bandwidth $h^{*}$ and choose $h(\mathbf{x}_{n})=h^{*}/P(\mathbf{x}_{n})$ for the kernel of order $p=2$ like Gaussian kernel. More details about the choice of kernel and bandwidth have been discussed in \cite{bugeau2007bandwidth}\cite{scott2015multivariate}\cite{sheather2004density}
.

In \cite{herrmann2017eigenstates} 
the author proposed a recursion relation to find proper bandwidth efficiently related to the priori estimator of density. As a grid method, the grid points $\mathbf{X}=\{\mathbf{x}_{1},\dots ,\mathbf{x}_{N}\}$ are distributed according to density $P(\mathbf{q})$ which is also estimated by a priori density $\widetilde{P}$ with respect to $\mathbf{X}$. For the requirement extended to multidimension according to \cite{herrmann2017eigenstates}
\begin{equation}\label{5}
\frac{1}{N}\sum_{n=1}^{N}\psi(\mathbf{x}_{n})\approx\int d\mathbf{q}\widetilde{P}(\mathbf{q})\psi(\mathbf{q})\approx\sum_{n=1}^{N}\int_{\text{Cell}_{n}}d\mathbf{q}\widetilde{P}(\mathbf{x}_{n})\psi(\mathbf{x}_{n})=\sum_{n=1}^{N}\lvert\text{Cell}_{n}\rvert\widetilde{P}(\mathbf{x}_{n})\psi(\mathbf{x}_{n}),
\end{equation}
for arbitrary smooth enough function $\psi(\mathbf{q})$ where $\lvert\text{Cell}_{n}\rvert$ denotes the volume of the cell containing $\mathbf{x}_{n}$ using Voronoi triangulation. The priori density is
\begin{equation}\label{6}
\widetilde{P}(\mathbf{x}_{n})=\frac{1}{N\lvert\text{Cell}_{n}\rvert}.
\end{equation}
The recursion relation is given by
\begin{equation}\label{7}
h_{n}\leftarrow h_{n}\frac{P_{N}(\mathbf{x}_{n};\mathbf{X})}{\widetilde{P}(\mathbf{x}_{n})}.
\end{equation}
In later computation more details will be discussed.
\section{Asymptotic smooth model}
Due to the infinity at origin of coulomb potential the MIW dynamical algorithm failed in evolution from given initial states to expected stationary states. The worlds or trajectories near origin will accelerate to origin till too close to crash then go to chaos. 
If the velocity Verlet algorithm is insisted in MIW dynamical algorithm, the classical process makes it impossible to solve the stationary states problem about the potential without local minimum. We turn to find some asymptotic smooth form of such potential with local minimum. The ``asymptotic'' means the tested potential will converge to target standard potential when some parameters tend to infinity. The ``smooth'' means the tested potential is continuous and differentiable at most domain. In practice, the ``asymptotic'' and ``smooth'' may not be guaranteed at particular points since necessary finite region and boundary condition in computation. In most simulations, few discontinue points distributed on boundary are tolerable in dealing with the infinity at origin. We expect that the results of MIW simulation by asymptotic smooth model approach the numerical simulation of standard quantum mechanics by matrix method as the parameters limits. 

One most used method in solution of the system with singular potential is cutting off which of the form (dimensionless unit $me^{2}/\hbar^{2}=1$ used in coulomb potential)
\begin{equation}
-\frac{1}{r}\rightarrow V_{a}(r)=-\frac{1}{r+a},\ \lim_{a\rightarrow0}V_{a}(r)=-\frac{1}{r}.
\end{equation}
This model has a local minimum at origin but the derivation is discontinue 


The smooth model of far field approximation for the coulomb potential has been studied in \cite{gonzalez2016smooth} 
using error function with the replacement
\begin{equation}\label{9}
-\frac{1}{r}\rightarrow V_{\mu}(r)=-c\ \text{exp}\left(-\alpha^{2}r^{2}\right)-\frac{\text{erf}(\mu r)}{r},\ \lim_{\mu\rightarrow\infty}V_{\mu}=-\frac{1}{r}
\end{equation}
where $c$ and $\alpha$ are linear in $\mu$ as the simplest choice. This model satisfies a local minimum $\lim_{r\rightarrow0}V_{\mu}(r)=-c-2\mu/\sqrt{\pi}$ and differentiable on all space. Actually only the error function term $-\text{erf}(\mu r)/r$ also satisfies these two constraints with local minimum $-2\mu/\sqrt{\pi}$ at origin. In \cite{gonzalez2016smooth} the author provided the comparison about more options. Error function can be replaced by
active function like hyperbolic tangent function or sigmoid function in dealing with singularity of coulomb potential at origin. 

The functions mentioned above can also be used in smooth the potential of the form with heaviside function. Consider a smooth model of finite depth well with width $2a$ and depth $L$
\begin{equation}\label{10}
V(x)=\left\{
\begin{aligned}
&0\ &\textrm{$-a<x<a$},\\
&L\ &\textrm{otherwise}
\end{aligned}\right.
\end{equation}
with the replacement
\begin{equation}\label{11}
V(x)\rightarrow V_{\nu}(x)=\frac{L}{2}\left[\text{erf}(\nu(x-a))-\text{erf}(\nu(x+a))+2\right],
\end{equation}
where $\nu$ is a flexible parameter to control the smoothness.\par
With these smooth models we expect to obtain some simulation by MIW dynamical algorithm and show the convergency when they tends to original models and the consistency between MIW approach and standard quantum mechanics.
\section{Simulation}
A standard MIW dynamical algorithm mainly contains two steps:
\begin{enumerate}
  \item Set velocities $\dot{\mathbf{X}}$ to zero after each iteration.
  \item Integrate $m\ddot{\mathbf{x}}_{n}(t)=-\nabla\left[V(\mathbf{q})+Q(\mathbf{q})\right]|_{\mathbf{q}=\mathbf{x}_{n}(t)}$ over a proper time interval $\Delta t$ and replace $\mathbf{x}_{n}(0)$ by $\mathbf{x}_{n}(\Delta t)$.
\end{enumerate}
After enough iterations the total energy should reduce to local minimum and worlds configuration tends to stationary configuration which would satisfy $\nabla\left[V(\mathbf{q})+Q(\mathbf{q})\right]|_{\mathbf{q}=\mathbf{x}_{n}(t)}=0$.


The selection of time interval is important in simulation. The simplest choice is fixed interval which requires the worlds configuration varying slowly enough. However, the too small interval is inefficient in the whole evolution because of the difference between the beginning and the end. Actually the evolution at the beginning allows larger interval since the initial configurations are incompact. An underlying problem occurs when we increase the number of worlds. For the efficient calculation we will restrict the region of initial worlds distribution which means more worlds will be closer. The vibration happened after some point near stationary state because of the fixed time interval. When the system approaches the stationary state, the fixed time interval seems too large to control the velocity which causes the over-evolution and vibration. When the system steps in some period we need control the interval to avoid the crossing of trajectories and keep the efficiency, hence lots of test needed in the fixed interval. There are lots of work about optimizing the verlet method \cite{huang1997adaptive}
\cite{hut1995building}
by reparameterization and variable time steps. In \cite{sturniolo2018computational} 
an adaptive time interval has been used which seems more efficient. 

\begin{figure}
  \centering
  \subfigure[]{
  \includegraphics[width=5cm]{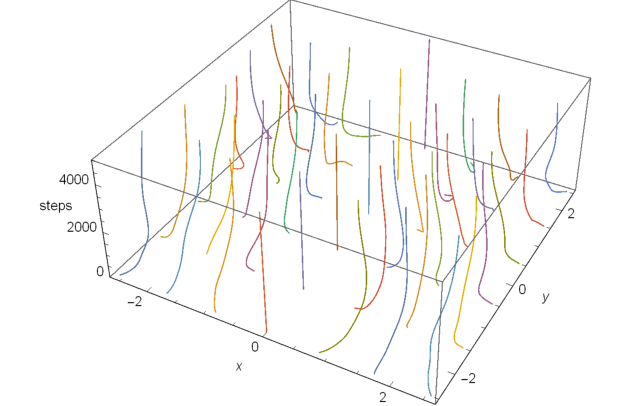}
  }
  \quad
  \subfigure[]{
  \includegraphics[width=5cm]{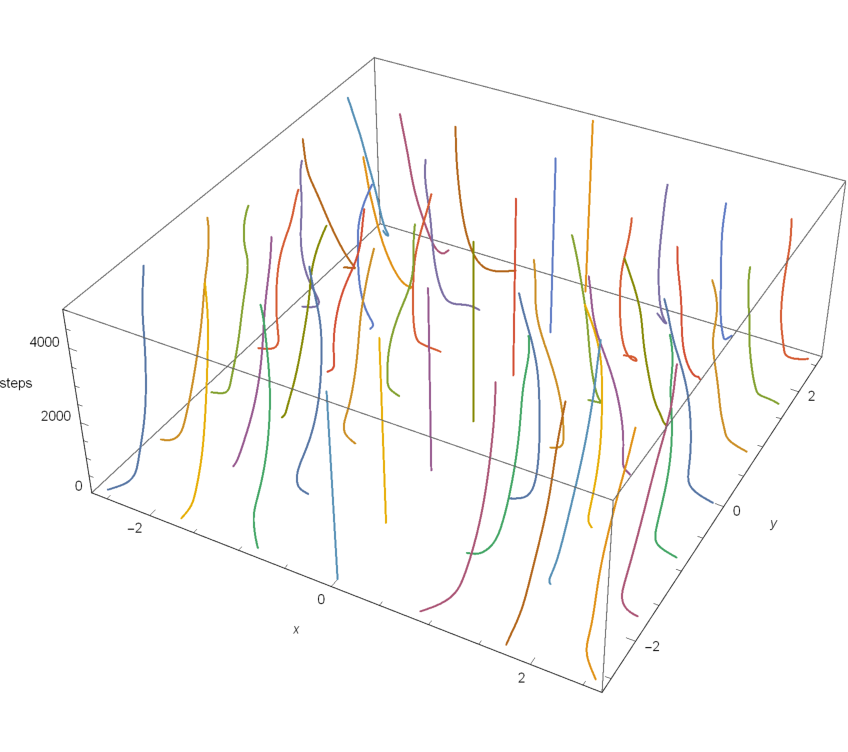}
  }
  \quad
  \subfigure[]{
  \includegraphics[width=5cm]{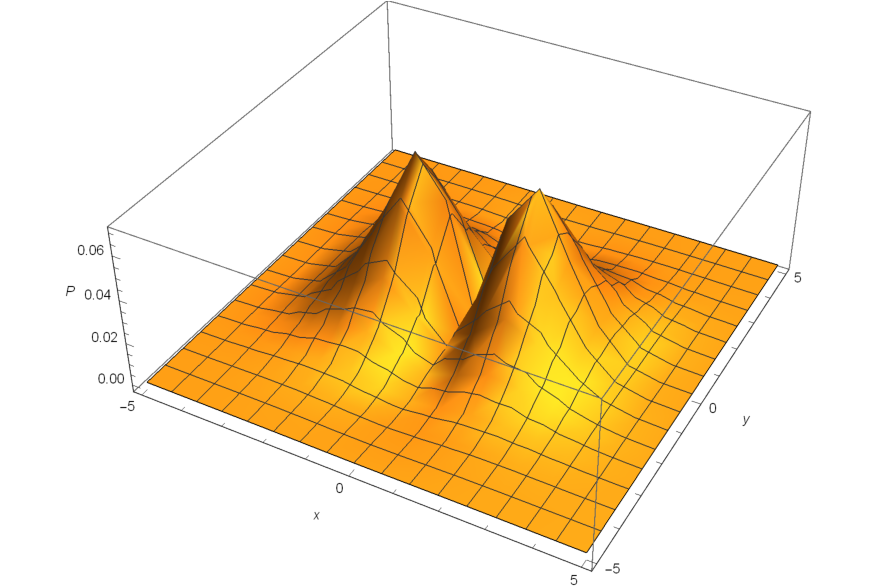}
  }
  \quad
  \subfigure[]{
  \includegraphics[width=5cm]{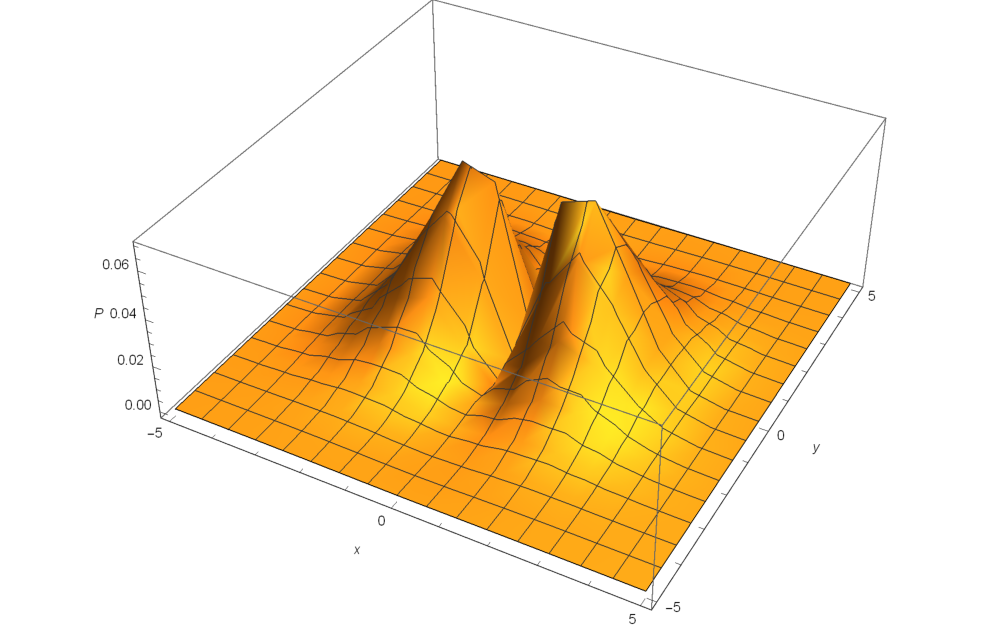}
  }
  \quad
  \subfigure[]{
  \includegraphics[width=5cm]{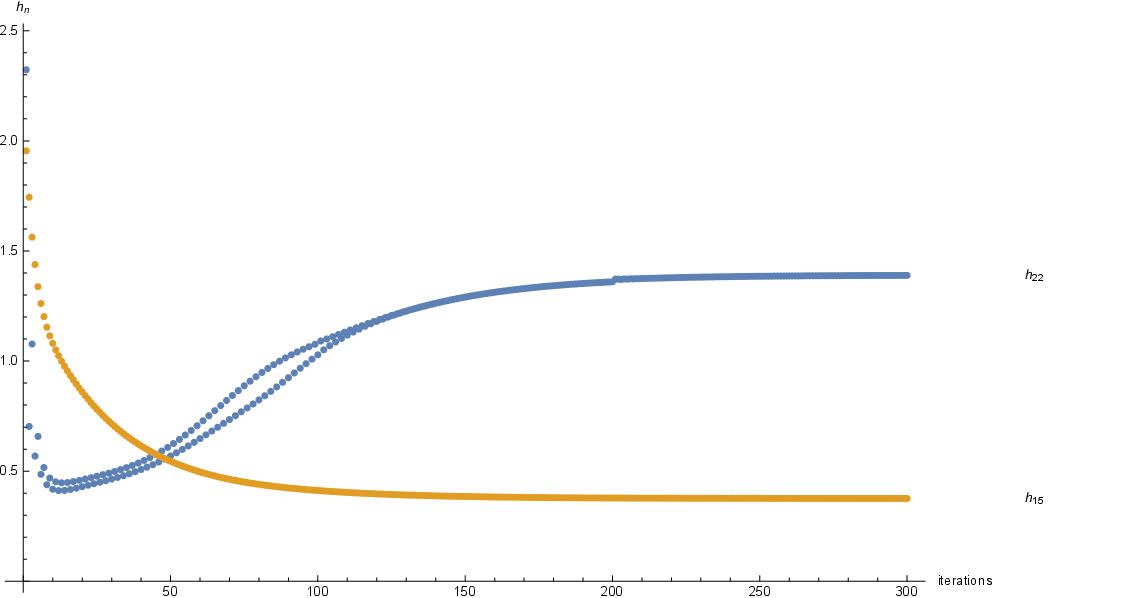}
  }
  \quad
  \subfigure[]{
  \includegraphics[width=5cm]{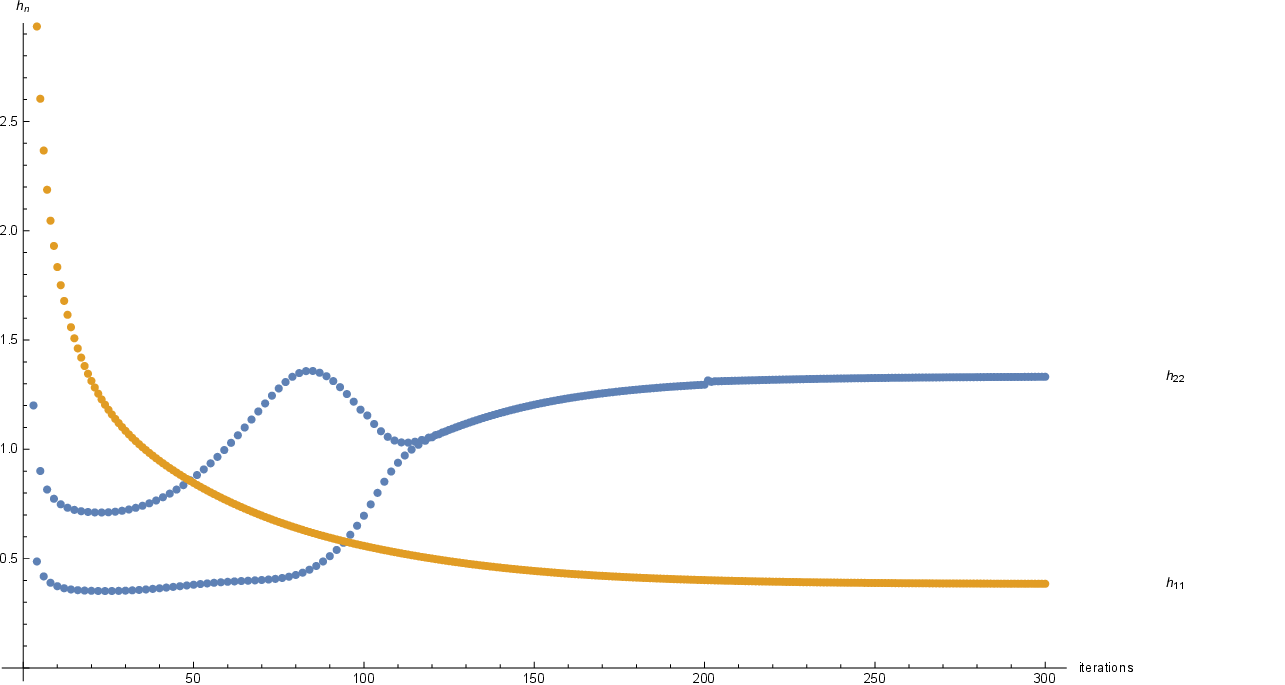}
  }
  \quad
  \subfigure[]{
  \includegraphics[width=5cm]{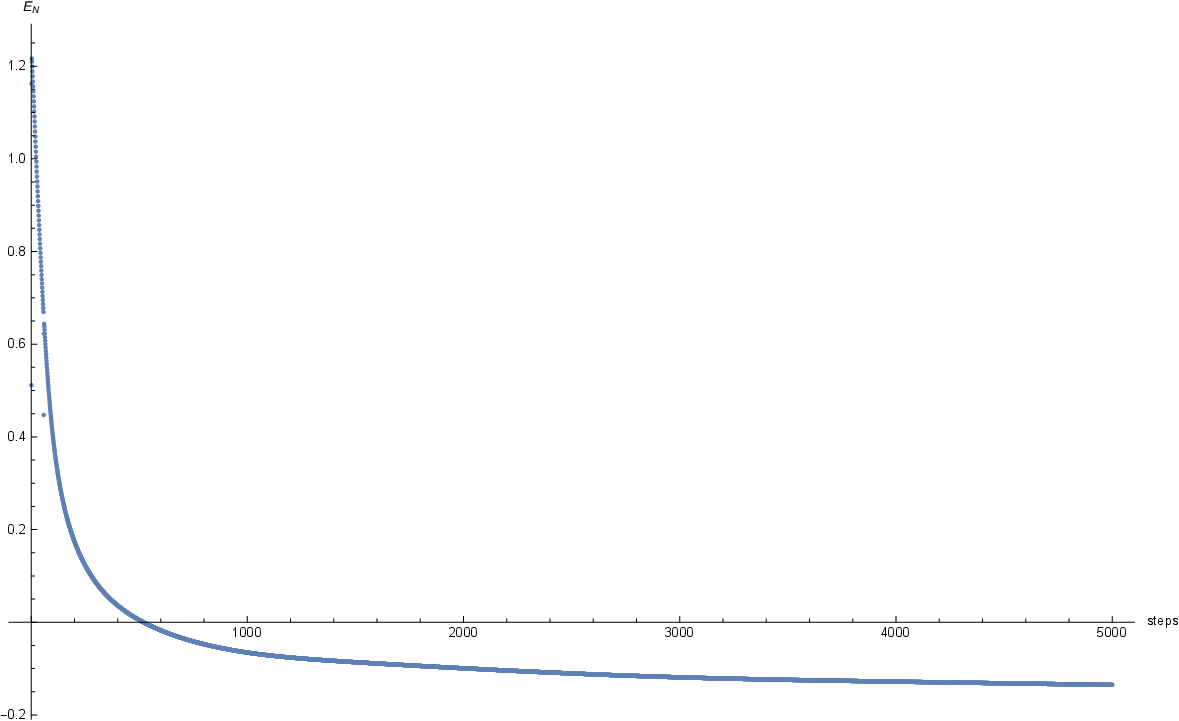}
  }
  \quad
  \subfigure[]{
  \includegraphics[width=5cm]{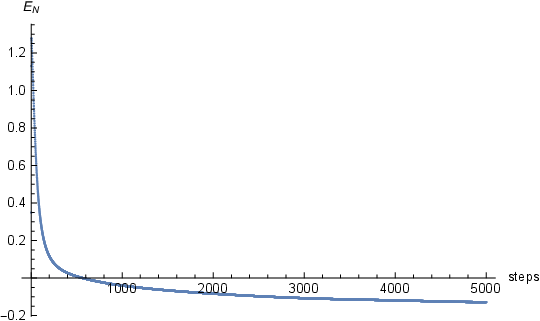}
  }
  \caption{Simulation results of excited states of error function model of 2-dimensional coulomb potential. The number of worlds in left column is $N=6\times7$, while $N=7\times7$ in right column. The models used here both are $V(x,y)=-\text{erf}(3\sqrt{x^{2}+y^{2}})/\sqrt{x^{2}+y^{2}}$. Some nodes are preset at $x=0$ with proper bandwidth conditions ($h_{20}=h_{23}=\infty$ in the former while $h_{23}=h_{25}=h_{27}\infty$ in the latter) which can be obtained after enough evolution. (a)(b): The evolution of worlds configuration of total 5000 steps in minimum effective region $3\times3$ square. (c)(d): The approximation of probability density function by kernel estimator. (e)(f): Some bandwidth varies after 300 iterations (100 iterations in each step) and converges within 3 steps. (g)(h): The energy of systems.}\label{fig1}
\end{figure}

Another choice to control the evolution is the recursion of bandwidth. At each time of iteration, we expect the estimator approaching the priori density which is fixed until next iteration. By the monotonicity of kernel function with respect to bandwidth $h$, we can obtain lots of convergent recursion of bandwidth. The recursion (7) proposed in \cite{herrmann2017eigenstates} 
repeats several times in each iteration with fixed grids and priori density in 1D ground state simulation. When it comes to excited states generalized recursion was proposed in \cite{herrmann2017eigenstates} 
to deal with nodes (priori density vanished at nodes). In 2D excited states simulation, we still follow the similar strategy in \cite{herrmann2017eigenstates} 
but make difference in dealing with bandwidth at nodes and node domain. We keep the minus in front of the kernel function of nodes since the failure to introduce the negative bandwidth in even kernel function. That means we need a new recursion for the bandwidth at nodes. Here we propose the recursion
\begin{equation}\label{12}
h_{n}\leftarrow h_{n}\frac{\sum_{i\in\text{nodes}}\frac{1}{h(\mathbf{x}_{i})^{2}}K\left(\frac{\mathbf{x}_{n}-\mathbf{x}_{i}}{h(\mathbf{x}_{i})}\right)}{\sum_{i\in\text{node domain}}\frac{1}{h(\mathbf{x}_{i})^{2}}K\left(\frac{\mathbf{x}_{n}-\mathbf{x}_{i}}{h(\mathbf{x}_{i})}\right)}.
\end{equation}
for the grid points at nodes, and keep the same recursion
\begin{equation}
h_{n}\leftarrow \frac{K(0)}{\widetilde{P}(\mathbf{x}_{n})-P_{N}(\mathbf{x}_{n};\mathbf{X})+h_{n}^{-1}K(0)}
\end{equation}
in \cite{herrmann2017eigenstates} 
for the grid points belong to the node domain. As shown in Fig.\ref{fig1} it works in simulation of 2D excited states.

Some boundary conditions are also necessary in simulation since we have to restrict the region of evolution if we consider the recursion of bandwidth by triangulation of configuration space. Though all worlds are equally weighted and special world is not allowed in MIW, we need some boundary worlds to represent the contribution of the worlds out of limited region. In the initial uniform grid cases of 2d coulomb potential we set 4 worlds in corner with fixed configuration and infinity bandwidth. 
These conditions can also be inferred by the recursion of bandwidth if we don't fix them in advance. More restrictions can be inferred by the recursion of bandwidth in the simulation of 2D excited states like infinity bandwidth at several nodes grid points. The restricted region can not be arbitrary small since not all worlds out of the region vary slowly enough to be represented by the same boundary conditions. Thus we propose the minimum effective region where the dynamical evolution converges to stationary states with proper boundary conditions. The minimum effective region limits the parameters in simulation for the machine precision limit in calculation. In most situation we can obtain this region according to the potential and target density by cutting off the region nearly vanishing. In practice we can consider the symmetry of distribution of worlds and simplify the computation. For example in a square region for the ground state, we actually only calculate any quarter,while in circle region calculation is only needed for the worlds along any radius. The symmetry also contributes in excited states since we need preset several nodes worlds as shown in Fig.\ref{fig1}.
Though the restricted region may cause some difference that is ignorable, it actually ensure the efficiency of evolution of systems.

\begin{figure}
  \centering
  \subfigure[]{
  \includegraphics[width=5.5cm]{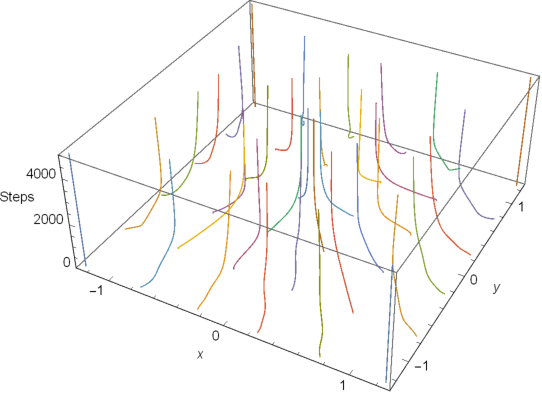}
  }
  \quad
  \subfigure[]{
  \includegraphics[width=5.5cm]{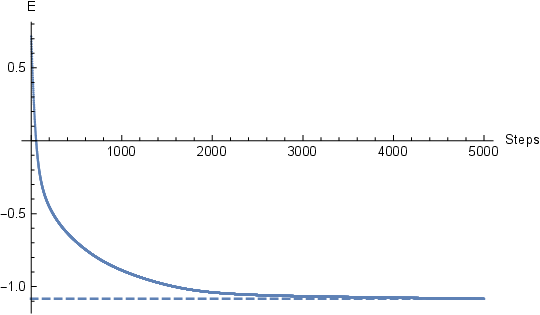}
  }
  \quad
  \subfigure[]{
  \includegraphics[width=5.5cm]{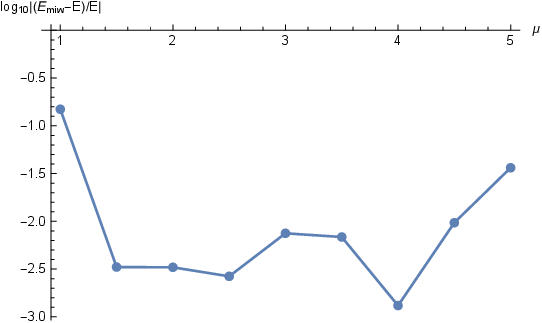}
  }
  \quad
  \subfigure[]{
  \includegraphics[width=5.5cm]{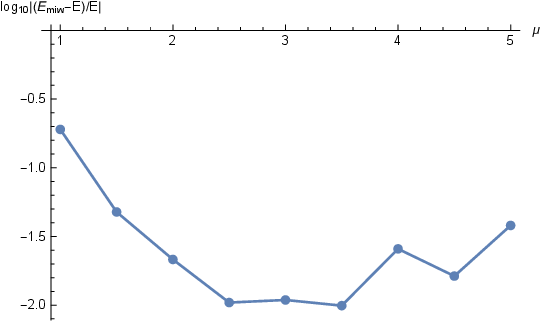}
  }
  \caption{Ground-state simulation results of smooth models of 2-dimensional coulomb potential. (a): The evolution of $N=6\times6$ worlds configuration of smooth model $V(x,y)=-\text{erf}(4\sqrt{x^{2}+y^{2}})/\sqrt{x^{2}+y^{2}}$ with $\Delta t=7.5\times10^{-2}$ and total 5000 steps in minimum effective region $3\times3$ square. (b): The energy $E_{\text{miw}}$ of the system in (a) (solid line) and ground state energy $E$ of the same system (dashed line) by matrix Numerov algorithm with grids of $59\times59$. (c): The energy error varies with parameter $\mu$ in error function model $V(x,y)=-\text{erf}(\mu\sqrt{x^{2}+y^{2}})/\sqrt{x^{2}+y^{2}}$. (d): The energy error varies with parameter $\mu$ in error function model $V(x,y)=-\text{tanh}(\mu\sqrt{x^{2}+y^{2}})/\sqrt{x^{2}+y^{2}}$.}\label{fig2}
\end{figure}

In the 2d coulomb potential simulation we tested error function model and hyperbolic tangent function model as shown in Fig.\ref{fig2}.
Within the scope of computation of personal computer, we compare the results of ground state obtained through MIW dynamical algorithm and standard matrix Numerov algorithm. The results of excited states shown in Fig.\ref{fig1} contain different nodes conditions that even nodes on the left and odd nodes on the right. With proper nodes conditions we can obtain the similar excited states to standard quantum mechanics. As we expected MIW approach shows the convergency and consistency to standard quantum theory. 



\begin{figure}
  \centering
  \subfigure[]{
  \includegraphics[width=5.5cm]{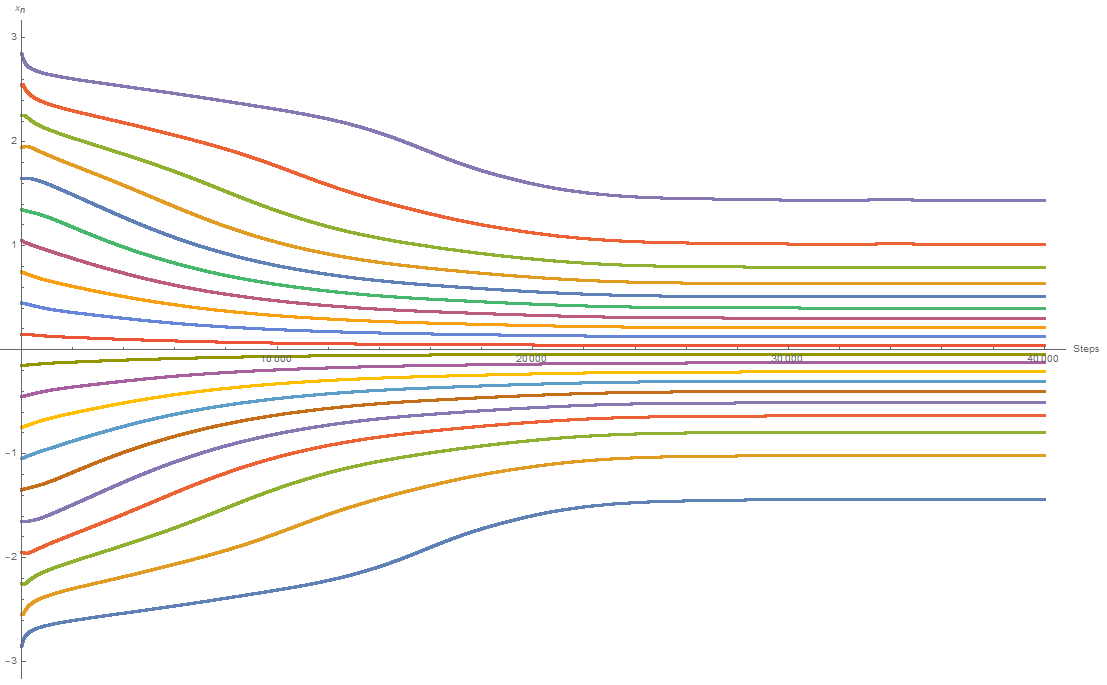}
  }
  \quad
  \subfigure[]{
  \includegraphics[width=5.5cm]{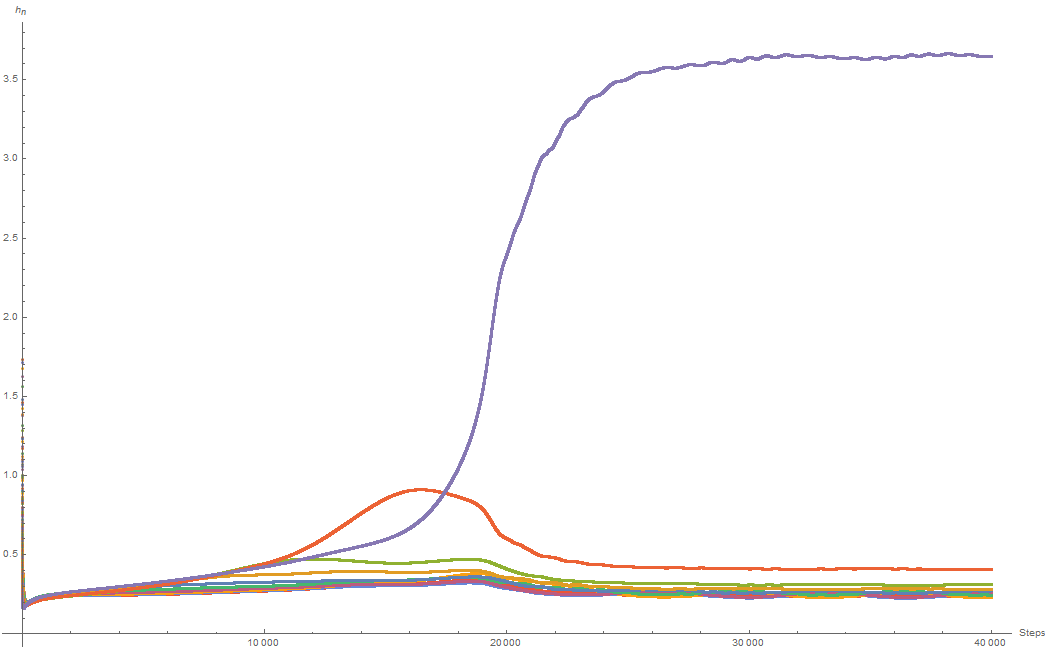}
  }
  \quad
  \subfigure[]{
  \includegraphics[width=5.5cm]{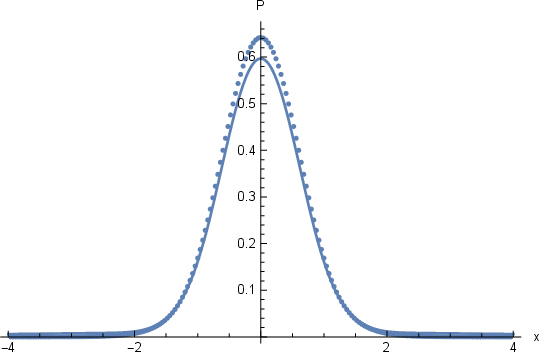}
  }
  \quad
  \subfigure[]{
  \includegraphics[width=5.5cm]{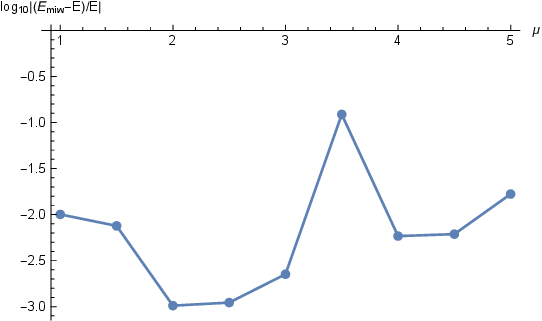}
  }
  \caption{Simulation results of smooth models of 1-dimensional finite trap potential. (a): The evolution of N=20 worlds configuration of smooth model $V(x)=\text{erf}(2(x-1))-\text{erf}(2(x+1))+2$ with $\Delta t=5\times10^{-2}$ and total 40000 steps in minimum effective region $\left[-4,4\right]$. (b): The bandwidth $h_{n}$ of the system in (a). (c): The density of ground states of system in (a) (solid line) and the solution from matrix Numerov algorithm with grids of $100$ (plots). (d): The energy error varies with parameter $\nu$ in error function model $V_{\nu}(x)=\left[\text{erf}(\nu(x-1))-\text{erf}(\nu(x+1))+2\right]$.}\label{fig3}
\end{figure}

In the 1d finite depth well smooth model of $L=2,\ a=1$, we obtain several expected consequences as shown in Fig.\ref{fig3}. 
Within the scope of computation of personal computer, we compare the results of ground state obtained through MIW dynamical algorithm and standard matrix Numerov algorithm. As we expected MIW approach shows the convergency and consistency to the standard quantum theory. The vibration occurs severely with the increase of $\nu$. With recursion of bandwidth the vibration can be suppressed but unavoidable. The large $\nu$ means sharp boundary at $x=-1$ and $x=1$. The worlds configuration near both boundary varies too fast while the recursion of bandwidth cannot reach the fitting value in time with the fixed time interval. Once the worlds evolute into the well region, the algorithm guarantee the equilibrium after enough evolution. The final state of system actually tends to the solution of standard quantum mechanics.

These simulation of different asymptotic models shows the consistency with traditional matrix Numerov method in solving the stationary states problem. With energy-decreasing algorithm and local-minimum models MIW approach can deal with most systems. However the precision of simulation is limited by the computer performance and software. In this article we used personal computer and Mathematica to calculate. The simulation became unreliable when the asymptotic parameters over particular value for the accumulated error after long time calculation. For example, the 2D coulomb potential simulation will evaluate to chaos when the $\mu$ in error function model larger than 5 or the number of grids larger than 36 in $3\times3$ square region.

\section{Discussion}
The MIW dynamical algorithm works as we expected in searching the stationary states for some typical systems by energy decreasing. It provides an iteration method to approximate the standard quantum solution of the given systems. As a asymptotic theory based on de Broglie-Bohm mechanics, it is constructed on density functions not wave functions and preserves the determinism. With proper density approximation and parameters iteration, it actually approachs the standard quantum mechanics. The dynamic algorithm can deal with larger number of worlds than the exact solution which is limited by the increasing complexity of the numerical solution. It's also convenient to be extended to multi-dimensional cases with triangulation and preset nodes. However the finite worlds and finite region limit the accuracy of simulation. We expect to theoretically obtain some asymptotic results like $E_{N,\nu}$ to prove $\lim_{N,\nu\rightarrow\infty}E_{N,\nu}=E_{n}$ which is the energy of nth excited states of the given system, but the iteration method makes the proceed difficult. Our strategy is the statistics analysis through enough simulation results that is time consuming.

Comparing to the adaptive quantum monte carlo approach based on Bohm description of quantum mechanics and supersymmetric quantum mechanics in dealing with nodes problems \cite{pladevall2019applied}
, the MIW dynamical algorithm performs better in extension to higher dimensions. But the strategy of presetting nodes makes it hard to solve the high excited states with complex nodes distribution. Without symmetry and presetting nodes, MIW dynamic is more disappointing in the system of potential without local minimum where the energy decreasing fails in evolution to stationary states.

The simulation of multi-dimension systems can be used in some multi-particle systems. For example, the 1D coupled harmonic oscillator with interacting $-Jx_{1}x_{2}$ \cite{park2018dynamics} 
can be viewed as 2D harmonic oscillator by diagonalization of Hamiltonian \cite{makarov2018coupled}
. The density matrix of nth and mth states can be approximation by higher dimension density function which makes it possible to compute the R$\acute{\text{e}}$nyi and Neumann entropies. However, the entanglement part of MIW still remain defective for the lack of phase of wave function.

The MIW approach provides two ways to solve a given system. The analytically exact solution shows the accuracy of MIW in approaching standard quantum mechanics but refers to the solution by standard quantum mechanics which makes it difficult to solve independently. The dynamical algorithm can advance the system to stationary states within the Bohmian mechanics and cost less calculation but limit the available potential. Without wave function the MIW approach lacks enough discussion about the phenomenon related to phase in standard quantum mechanics. In the future we expect to perfect MIW approach by stepping back to Bohmian mechanics and the ontology of wave function.
\section*{Acknowledgements}
The research was supported by National Key R\&D Program of China under Grant No.2018YFB1601402-2.
\bibliography{one}
\bibliographystyle{plain}
\end{document}